\documentclass[prb,floats,aps,superscriptaddress,showpacs,twocolumn]{revtex4}
\usepackage{amsmath}
\usepackage{graphicx}
\newcommand{\be}{\begin{equation}}
\newcommand{\ee}{\end{equation}}
\newcommand{\ba}{\begin{eqnarray}}
\newcommand{\ea}{\end{eqnarray}}

\begin{document}

\title{Transport of a heated granular gas in a washboard potential}
\author{Giulio~Costantini}
\affiliation{Dipartimento di Fisica, Via Madonna delle Carceri,
68032 Camerino (MC), Italy}
\author{Fabio~Cecconi}
\affiliation{INFM Center for Statistical Mechanics and Complexity,  
P.le Aldo Moro 2, 00185 Rome (Italy) \\ 
and Institute for Complex Systems (CNR), 
Via dei Taurini 19, 00185 Rome (Italy).}
\author{Umberto Marini-Bettolo-Marconi}
\affiliation{Dipartimento di Fisica, Via Madonna delle Carceri,
68032 Camerino (MC), Italy}

\date{\today}

\begin{abstract}
We study numerically the motion of a one dimensional array of Brownian 
particles in a washboard potential, driven by an external stochastic force and 
interacting via short range repulsive forces.
In particular, we investigate the role of instantaneous elastic and inelastic 
collisions on the system dynamics and transport.
The system  displays a locked regime, where particles may move only via
activated processes and a running regime where particles drift
along the direction of the applied field.
By tuning the value of the friction parameter controlling
the Brownian motion we explore both the overdamped dynamics
and the underdamped dynamics. In the two regimes 
we considered the mobility and the diffusivity of the system as functions
of the tilt and other relevant control parameters such as,  
coefficient of restitution, particle size and total number of 
particles.  We find that, while in the overdamped regime, the results for 
the interacting systems present similarities with the known non-interacting 
case, in the underdamped regime, the inelastic collisions determine a rich 
variety of behaviors among which is an unexpected enhancement of the 
inelastic diffusion.

\end{abstract}
\pacs{05.40.-a,61.20.Gy.05.10.Gg}
\maketitle

\section{Introduction}

 A large class of phenomena in biology, chemistry, engineering 
and physics occurs via the transport of particles driven along periodic 
substrates by an external bias.
These phenomena include polymer diffusion at
interfaces~\cite{polimeri}, motion of fluxons in superconductors
containing a periodic arrangement of defects ~\cite{Nori1,Barone},
adsorption on crystal surfaces~\cite{adsorbate}, super-ionic
conduction~\cite{superionic}, motion of molecular motors along
microtubules~\cite{Prost}, granular flows on rough inclined
substrates ~\cite{Vicsek,vulpi,wolf,Henrique,Ancey} or in a narrow
pipe~\cite{pipe}.

The study of transport properties of granular systems represents
an open issue in Statistical Mechanics of considerable interest and 
difficulty in view of the continuous energy dissipation caused by
particle inelastic interactions~\cite{Farkas}.
In this perspective,
we consider a simple model consisting  of a granular fluid moving on
a tilted rough substrate, which may favor clustering and jamming behaviors.

 In our formulation, a granular system is a large number of particles 
(grains) colliding with one another and losing a little energy in each 
collision~\cite{Jaeger,General1,General3,General4,General5}.
If such a system is shaken to keep it in motion 
its dynamics resembles that of fluids as the grains move 
randomly.~\cite{Paolotti}
 
We carry out a comparative study of the behavior of three models:
the inelastic particle system (IPS), the elastic particle system (EPS), and  
finally the non interacting system or independent Brownian particles (IBP) 
to understand how the interactions influence the collective transport.  
 The absence of interactions between particles makes the basic  
phenomenology of the IBP well understood. It reduces, indeed,  
to the motion of a single particle in a force field 
resulting from the interplay between a constant force $F$, such as gravity or
electric field,  
a spatially periodic force, simulating  the presence of a rough substrate,
a viscous force accounting for the friction,
and a noise term, representing thermal fluctuations ~\cite{Risken}.
The scenario is the following:
particles diffuse with a bias in the direction of the steady force, $F$,
with an average velocity, $v_m$,
which is an increasing function of the applied load $F$.
However, when the load is below a critical value, $F_c$, and the
noise sufficiently small, the average velocity tends to zero.
In other words the particles remain
locked in the minima of the periodic potential. On the other hand, 
above the threshold, $F_c$,
the particles may travel from one minimum to the other along the
tilt direction. 
The critical value, $F_c$, depends on roughness, 
friction and temperature.  According to the value of the friction coefficient 
two different 
scenarios may be observed. In the so called overdamped limit,
$F_c$ occurs only when the tilt is such that the local minima 
completely disappear. In the opposite limit, {\em i.e.}, in the
underdamped regime, the particles may overcome many barriers
even for low values of $F$ at which the potential still displays 
local minima. In fact,
the particles may cross a barrier, provided their gain in potential
energy, to go from one minimum to the next,
exceeds the energy dissipated along this pathway.
In the underdamped limit  the exact value of $F_c$ depends on the 
temperature.

Two indicators may serve to characterize the
collective transport properties of the system, the mobility, $\mu$,
and the diffusion constant, $D$.
The former relates the average velocity to the tilt $F$ according to  
$ \langle v \rangle = \mu(F) F$. The latter measures the average
spreading of particles:
\begin{equation}
R(t) = \langle [x(t) - \langle x_{CM}(t) \rangle]^2 \rangle\;.
\label{diffusion}
\end{equation}
 The average is meant over the thermal noise realizations. 
In all the cases where $R(t)$ asymptotically grows linearly in time,
we can identify the normal diffusion constant $D$ defined as the slope of the
law
$$
R(t) \sim 2D t
$$
that generally is computed numerically through simulations of the system
evolution.
The study of the dependence of $\mu$ and $D$ on $F$ and 
other system parameter is necessary to determine
the efficiency of the transport, i.e. the ratio between drift 
and spreading of a swarm of particles.

A detailed theory capable of describing 
the behavior of the IBP for arbitrary values of the damping 
has been developed by Risken~\cite{Risken,Coffey} who derived an
analytical expression for the mobility, $\mu$, 
as a function of noise level, tilting force and substrate 
characteristics.
A closed formula for the diffusion coefficient exists only for 
the overdamped situation~\cite{Reimann}.
It predicts the presence of a striking
enhancement of $D$ near a threshold, $F_3$, separating
the locked from the running phase and corresponding to the
disappearance of the local minima of the potential.

On the other hand, 
in the  underdamped regime there is no analytic expression for $D$,
but the enhancement of the diffusion coefficient, 
occurs at a lower threshold, $F_2$~\cite{Risken,Costantini}
due to the effect of inertia,
and the phenomenon remains qualitatively similar.

The goal of the present paper is to consider the effects of 
interactions among particles on the transport properties in such systems.  
Specifically, we will address the following question: how do interactions 
occurring via elastic/inelastic hard-core collisions affect the mobility 
and the diffusion of an assembly of particles? This issue is particular 
relevant in one dimension where
a repulsive hard-core interaction inhibits the particles to pass each 
other; this constraint is known to influence dramatically the dynamics of a 
group of particles. As an example, we can mention the anomalous self-diffusion 
in single file systems \cite{Harris,Kollmann,Taloni}.  

We focus our dynamical approach only on exploring the effects 
of repulsive interactions via impulsive contact collisions.  
We consider two possibilities, 
energy conserving collisions and dissipative inelastic collisions.
Both situations are still largely unexplored in systems with washboard 
potentials 
and display non-trivial
behaviors, as we shall illustrate.
For instance, the mutual repulsion between the particles induces
dynamical correlations which may favor or hinder their motions:
it can promote their exit from a potential well via
energetic collisions, or, on the contrary, it can prevent
a jump to a given site when this is too crowded.
In addition, the granular temperature, 
defined as the average kinetic energy per particle,
in inelastic systems is in general lower than the corresponding temperature
of elastic systems. Therefore, 
the transport, occurring via thermally activated processes across
substrate barriers, is expected to be less efficient for inelastic 
particles~\cite{Biwell}. 
Then few questions can be addressed. Does the diffusivity $D$ present
an enhancement analogous to that of non interacting systems?
Does the threshold locked-to-running occur at larger values of the tilt?
In the following, we answer the questions by examining a variety of 
situations and analyzing the overdamped and the underdamped 
regimes separately.

The paper is organized as follows. In section II, we present the
model and discuss the contact elastic/inelastic interaction with
their main implication in the system dynamics. In Section III, we 
analyze the results from the molecular dynamics (MD) simulations of the
over-damped  regime. In section IV, we
summarize the corresponding results for the under-damped
regime discussing the salient differences with the over-damped case.
Finally, conclusions are drawn in section V.

\section{Model}
We consider a randomly driven granular gas, already introduced in 
previous 
works~\cite{CDBP}, composed of $N$ impenetrable 
hard-rods, of mass $m$ 
and size $\sigma$, moving on a line of length $L$ in the presence of an 
external washboard potential.
The overall dynamics of this gas is described by the Langevin equation 
for each rod
\begin{equation}
m \frac{d^2 x_i(t)}{d t^2} =
- m \gamma\frac{d x_i(t)}{d t} 
+ \xi_i(t) + \sum_{j\neq i} f_{ij}
- \frac{d \Psi[x_i(t)]}{d x_i}
+ F 
\label{kramers2}
\end{equation}
that embodies four types of physical phenomena: friction with the 
surroundings, random accelerations due to external driving, inelastic 
collisions among the particles and external time independent, but spatially
varying, force.
We model the first two effects by means of a stochastic bath  
with a viscous friction $-m\gamma \dot{x}_i$ and 
Gaussian random force, with zero average and covariance 
\begin{equation}
\langle \xi_i(t) \xi_j(t') \rangle = 2 m \gamma
T\delta_{ij} \delta(t-t')
\label{variance}
\end{equation}
satisfying a fluctuation-dissipation relation,
with $T$ proportional to the intensity of the stochastic 
driving~\cite{Pagnani}. 
The damping term fulfilling Einstein's relation renders the 
system stationary even in the absence of collisional dissipation and
physically can represent the friction between the particles and the
container. The interactions
among the particles, indicated formally 
by $\sum_j f_{ij}$ in Eq. (\ref{kramers2}),
amount to simple impulsive forces acting on the rod $i$ due to the 
collisions with the neighboring particles 
$j$~\cite{Kadanoff,Sela,Mackintosh,Mcnamara}.  
Thus a rod $i$ performs, under the influence of the bath and collisions only, 
independent Brownian trajectories, unless it gets in contact 
with particle $j$, i.e.
$|x_{j}-x_i| = \sigma$, at which point the velocities of the colliding pair 
$(i,j)$ change instantaneously according to the inelastic rule
$$
v_i' = v_i - \frac{1+\alpha}{2}(v_i - v_j)
$$
$$
v_j' = v_j + \frac{1+\alpha}{2}(v_i - v_j),
$$
the prime indicating post-collisional variables and $\alpha$
the coefficient of restitution, with $0\leq\alpha\leq 1$.

The total force, in Eq.~(\ref{kramers2}), on each particle 
is supplemented by terms 
representing the presence of a constant external bias, $F$, and 
a periodic substrate generating a potential $\Psi(x)$, with period $w$
and amplitude $V_0$,
\begin{equation}
\Psi(x)= -V_0 \cos(2\pi x /w).
\label{psi}
\end{equation}
A quantitative criterion to distinguish between an overdamped and 
an underdamped
regime is provided by the dimensionless parameter $\Gamma=\gamma/\omega_0$,
where 
$$
\omega_0=\sqrt{\frac{1}{m}\Big[\frac{d^2\Psi(x)}{dx^2}\Big]}_{min}
$$
is the oscillation frequency at the potential minimum.
If $\Gamma \gg 1$ the dynamics is overdamped, and in the opposite limit is 
underdamped.

As discussed by Borromeo and Marchesoni \cite{Borromeo,Borromeo2}, the present
model is equivalent to consider the particles in a traveling potential, 
i.e. a periodic potential
moving at speed $c$, which is related to the tilt force by the
relation $F = m\gamma c$.
Interestingly for speeds 
lower than a certain threshold the traveling wave has the
capability of dragging the particles, a mechanism known as
Stokes' drift \cite{Stokes,Stokes2}.

Thus, the system can equivalently describe the
physics of a one dimensional granular gas where
the periodic potential $\Psi(x)$ represents a series
of compartments separated by walls,
an experimental set-up recently employed to study the
clustering behavior of vibrated granular gases~\cite{Lohse2,Conti},
or the roughness of an inclined plane~\cite{wolf,vulpi} .

\section{Nearly overdamped regime}
We shall begin by considering the nearly overdamped regime 
($\Gamma\simeq 2.1$), whose study
is better understood because the system reaches
a steady state rapidly due to the large value of the friction.
 
Our MD simulations were carried out
by evolving an initial configuration, where the
particles were all located in the central well without
overlaps 
and their velocities were extracted 
from a Maxwell-Boltzmann distribution of temperature $T$.
Each run involved $N=256$ particles of size $\sigma=1$,
mass $m=1$, with three different values of the
coefficient of restitution, $\alpha=0.8$, $0.9$, and  $1.0$, respectively.
The substrate potential was characterized by wells of width 
$w = 400\sigma$ and height $V_0 = 9.0$, in units of $k_B T$. Finally,
the heat-bath temperature has been chosen to be $T=1.0$ and $k_B=1$.
The time is measured in units of $t_u=\sigma\sqrt{m/T}$.
The friction coefficient is $\gamma=2/t_u$.
The cyclic system studied has been taken to be of length $L = 10^8 w$  
so that it is virtually equivalent to a system with open boundary conditions.
In this condition of high dilution, 
the global system density $\rho = N/L$ is extremely 
low, however it is not meaningful parameter, rather it is the initial 
density profile, 
characterized by the number of particles in a well $\rho_w = N/w$, that has
a strong influence on the system behaviour. Indeed, at the beginning of the 
evolution, the system needs to be packed enough to reach a not 
negligible collision rate. Only in the later stages of the simulations,  
a crossover is observed toward the behaviour of rarefied gases.   

\begin{figure}[htb]
\includegraphics[clip=true,width=8.5cm, keepaspectratio,angle=0]
{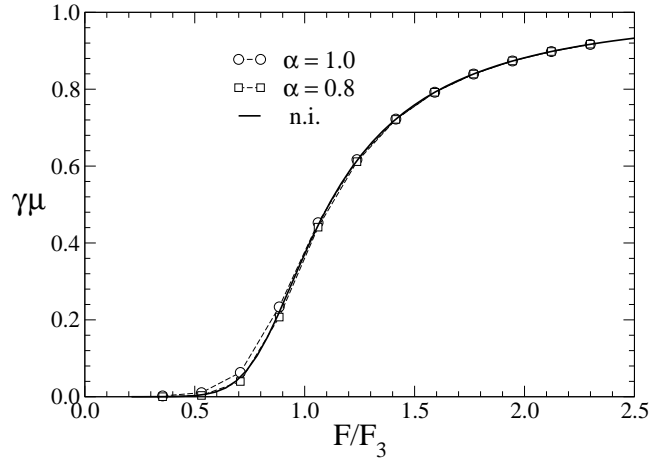}
\caption{Dimensionless mobility $\gamma \mu$ as a function of the
rescaled external force $F/F_3$ ($F_3=2\pi V_0/w$) for $\alpha =1.0$
(circles) and $\alpha =0.8$ (squares) for a system with 
$N=256$ particles, 
temperature $T=1$, friction $\gamma=2/t_u$, and potential amplitude 
$V_0=9.0$, $\sigma=1$, $w/\sigma=400$. The system evolution is 
simulated for $t_{max}=10000$ time units. 
The full line indicates the corresponding curve for the
non interacting particles obtained by the continued
fraction method \cite{Risken}.}
\label{fig:mob}
\end{figure}

\begin{figure}[htb]
\includegraphics[clip=true,width=8.5cm, keepaspectratio,angle=0]
{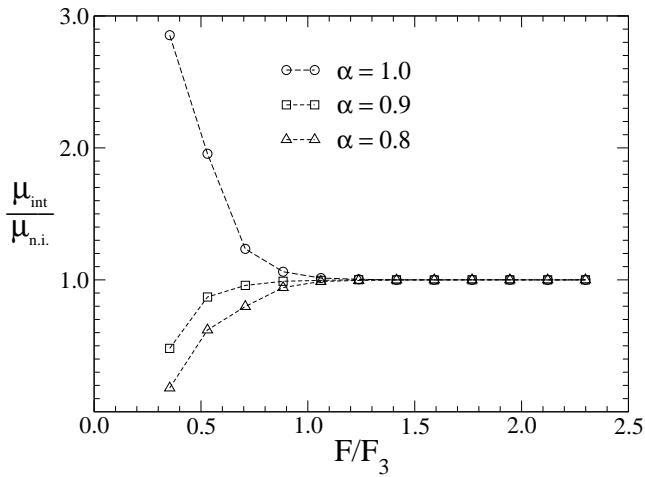}
\caption{The ratio between mobilities of interacting and
non interacting systems as a function of the rescaled external force
$F/F_3$ ($F_3=2\pi V_0/w$) for $\alpha=1.0$ (circles), $\alpha=0.9$
(squares) and $\alpha=0.8$ (triangles). The remaining parameters are
the same as in Fig.~\ref{fig:mob}.}
\label{fig:rapmob}
\end{figure}

At first sight, the MD simulations of the interacting system
display a mobility quite similar to that of
the non-interacting system: the curves $\gamma \mu$ 
versus $F$ for elastic and inelastic
hard-rod systems are shown in Fig.~\ref{fig:mob}. 
In the same figure, we also plot
the corresponding quantity relative to the IBP
as calculated numerically by means of the
continued fraction method~\cite{Risken}.
 However, a closer inspection reveals that differences do exist 
between the IBP and the interacting systems with different 
inelasticity $\alpha$.
Figure~\ref{fig:rapmob} shows the ratio of the 
mobilities of interacting systems to the mobility of the IBP.   
This ratio, for small values 
of the load, $F\ll F_3=2\pi V_0/w$, can be significantly different 
from $1$, 
and more specifically the mobility of the EPS 
exceeds the mobility of the IBP, whereas the mobility of the IPS 
is lower. The reason for these differences can be found in the  
fact that contact interactions may change drastically, 
with respect to the IBP, the time that particles spend in a given 
potential minimum.

In more detail, these behaviors can be explained by recalling that
two  competing mechanisms contribute to the system mobility: 
the excluded volume and the inelasticity. 
The first leads to an effective reduction of the
barrier height and favors the escape from the wells, thus basically
increasing the mobility. The second, instead, tends to decrease 
the average
kinetic energy, rendering longer, on average, the time
spent by the particles in a given well.
For low values of the forcing field $F$, the mutual repulsion
dominates and thus we observe a larger mobility
of the EPS with respect to the IBP at the same values of $F$ and $T$.
However, as we switch the inelasticity on, the mobility of the IPS
can become lower than that corresponding to the IBP.
For $F\gtrsim F_3$, $\mu$ turns out to be the same for all systems
because in this regime only the forcing field $F$ matters. In fact,
the absence of minima in the total potential reduces the
influence of the mutual interactions so the particles follow 
coherently the strong effect of the drift.

 The role of the excluded volume can be highlighted by monitoring 
the trajectories of some specific tagged particles
whose dynamics can be very representative of the evolution of the 
whole system. More specifically, we consider
the two extremal particles, 
$1$ and $N$ (having labeled the rods along the drift direction from 
$1$ to $N$) and the central particle $i=N/2$. 

\begin{figure*}
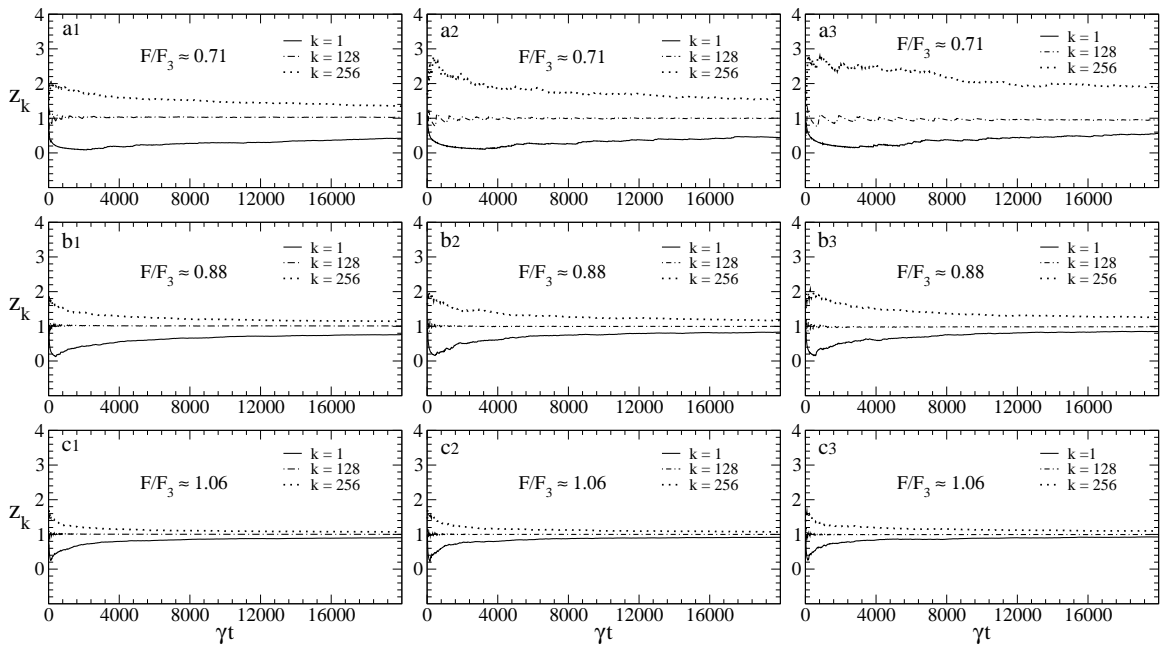

\includegraphics[clip=true,height=8.5cm, keepaspectratio]
{fig3a.eps}
\includegraphics[clip=true,height=8.5cm, keepaspectratio]
{fig3b.eps}
\includegraphics[clip=true,height=8.5cm, keepaspectratio]
{fig3c.eps}
\caption{Evolution of the normalized displacements 
from the initial position 
(see Eq.~(\ref{eq:tagged})) of the trajectories referring to tagged 
particles $k=1$ (full line), $k=128$ (dot-dashed) and $k=256$ 
(dots). The normalization is performed with respect to the 
displacement of the center of mass from its initial coordinate.
The panels a1,a2 and a3 refer to rescaled forces $F/F_3\simeq 0.71$, 
b1,b2 and b3 to $F/F_3\simeq 0.88$ and finally c1,c2 and c3 to 
$F/F_3\simeq 1.06$. 
The system contains $N=256$ particles, the coefficients of the restitution 
are $\alpha =1.0$ (in plots a1, b1, c1), $\alpha =0.9$ (in plots a2, b2, c2)
and $\alpha =0.8$ (in plots a3, b3, c3). Data are averaged over 
$10$ independent realizations and the remaining
parameters are the same as in Fig.~\ref{fig:mob}.}
\label{fig:raptraj}
\end{figure*}

The first particle is, on average, 
braked by the collisions of the preceding 
particles, whereas the $N$-th particle is pushed ahead by the pressure of 
those behind. Their overall 
behavior is expected to be qualitatively rather different from that 
of the central particle.
 In figure~\ref{fig:raptraj}, we display, for  
$\alpha=1.0,0.9$ and $0.8$, the evolution of the mean 
displacement of these three tracer particles from 
their initial position. Each displacement is normalized with respect to 
the displacement of the center of mass of the whole system   
from its initial position:
\begin{equation}
 z_k(t) =  \frac{\langle |x_k(t) - x_k(0)| \rangle }
{\langle |x_{CM}(t) - x_{CM}(0)| \rangle}
\label{eq:tagged}
\end{equation}
with $k=1, N/2 $ and $N$.  
All the averages are performed over $20$ independent runs.

We consider, first, the EPS for three values
of the tilt $F$, first column of Fig.~\ref{fig:raptraj}.
For all values of $F$, the central particle moves with velocity
very close to the center of mass velocity $v_{CM}$ as shown 
by the fact that $z_{N/2}$ stays almost pinned around the value $1$.
The velocities of the extremal particles, instead, 
are very different from $v_{CM}$ only during an    
initial transient when the system remains compact. Indeed, 
it is clear that, in a system not yet too diluted, 
the motion of the first particle is frequently hindered by the 
others,  while the motion of the last is favored.
Asymptotically, the differences in the motion of rods $1$, $N/2$ and $N$ 
become less evident because the interactions become less effective as 
the packet spreads over and over. Notice, also, the asymmetry 
between the first and the last particle.

\begin{figure}[htb]
\includegraphics[clip=true,width=8.5cm, keepaspectratio]
{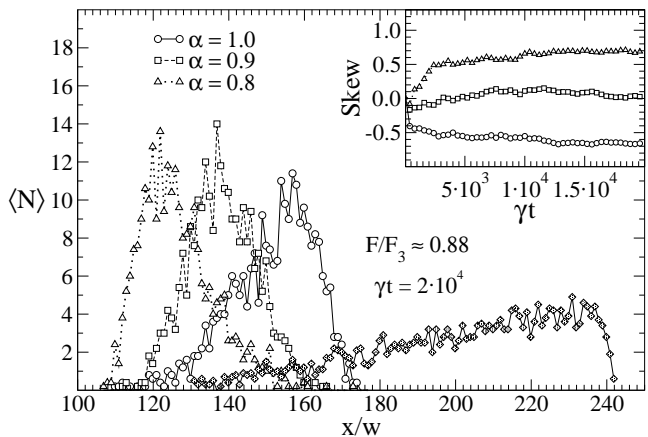}
\caption{Averaged number of particles, for different coefficients
of restitution, as a function the dimensionless quantity $x/w$ with
$F/F_3\simeq 0.88$ and $\gamma t=2\cdot 10^4$. The circles correspond
to the elastic case ($\alpha=1.0$) with $\sigma=1$, while 
the squares and triangles to $\alpha=0.9$ and $\alpha=0.8$,
respectively. The diamonds correspond, instead, to the elastic particles of
size $10\sigma$. The distributions extend only over few hundred  
wells, while the system size is much larger $L=10^8 w$. 
Inset: The trend of the skewness as a function of
$\gamma t$ in the same cases. The other parameters are the same as in
Fig.~\ref{fig:mob}.}
\label{fignx}
\end{figure}

The major change observed when the inelasticity is turned on ($\alpha<1$) 
is that 
the velocity of the last particle displays a pronounced
deviation from $v_{CM}$. 
These different behaviors are also evident by an inspection of the 
shapes of the instantaneous coarse grained distributions of particle 
positions (Fig.~\ref{fignx}), 
\begin{equation}
N(x,t) = \int_{x-w/2}^{x+w/2} dy \rho(y,t)
\label{eq:rho-cg}
\end{equation}
computed, in the simulations, by binning the number of particles to 
widths of the size of a single potential well. 

In the EPS, the asymmetry of the packet is determined
only by the excluded volume effects and, when the  diameter 
of the rods
increases from $\sigma$ to $10 \sigma$, both asymmetry and 
drift velocity become 
more pronounced.

 The dissipation, favoring clustering,  
makes the packets of the IPS more compact and, in addition, changes the 
direction of the asymmetry, as seen 
for $\alpha=0.9$ and $\alpha=0.8$, in Fig.~\ref{fignx}.
This effect can be quantified by means of the skewness of the particle
distribution around their center of mass,
\begin{equation}
\mbox{Skew}(x_1,...,x_N)
=\frac{1}{N}\sum_{j=1}^N\Big(\frac{x_j-x_{CM}}{\mbox{rmsd}}\Big)^3
\end{equation}
where $\mbox{rmsd}(x_1,...,x_N)$ is the standard deviation and $x_{CM}$ the
center of mass position.
Positive values of $\mbox{Skew}$ entail distributions with an asymmetric
shape extending out toward more positive tails. 
Negative values are associated with distributions extending
out toward more negative tails~\cite{Numericalrecipes}.
In the inset of Fig.~\ref{fignx}, we plot the
parameter $Skew$ as a function of time for three systems
with different inelasticities. 

\begin{figure}[htb]
\includegraphics[clip=true,width=8.5cm, keepaspectratio]
{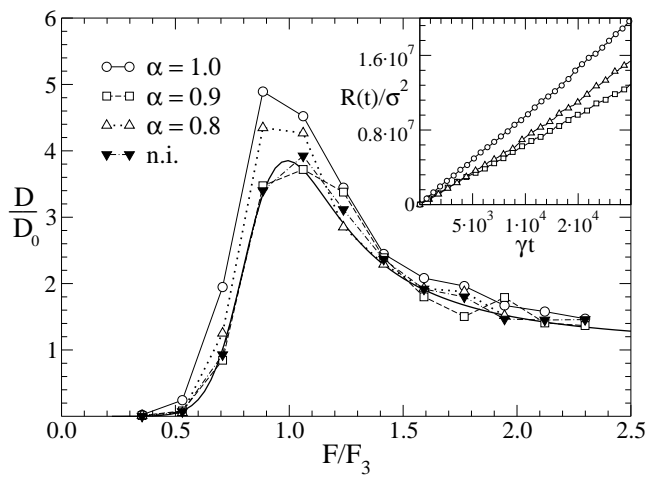}
\caption{The rescaled coefficient of diffusion $D/D_0$, ($D_0 = T/m\gamma$) 
as a function of $F/F_3$. Each point is the 
result of an average over $10$ independent runs. The circles correspond to the
elastic case ($\alpha=1.0$), while the squares and triangles to
$\alpha=0.9$ and $\alpha=0.8$, respectively.
The full line represents the theoretical result obtained by 
Reimann et al.~\cite{Reimann}. 
Inset: standard diffusion of the 
system rescaled spreading, $R(t)/\sigma^2$, [see Eq.~(\ref{diffusion})] 
with 
respect to the dimensionless time $\gamma t$ 
in the same cases for $F/F_3\simeq 0.88$. 
The other parameters are the same as in 
Fig.~\ref{fig:mob}.}
\label{fig:Diff-sv}
\end{figure}

Let us consider, now, the quantity 
\begin{equation}
D = \lim_{t\to \infty}\frac{1}{2 t} R(t)
\label{diffusion2}
\end{equation}
In the IBP, $D$ is constant and corresponds to the diffusion coefficient.
We have found numerically that such a behavior persists both
in the EPS and IPS for all values of $\alpha$ we explored,
as clearly indicated by the linear growth of $R(t)$ in the inset of 
Fig.~\ref{fig:Diff-sv}. 

The interactions change quantitatively the dependence
of $D$ on $F$, in fact,    
as $F$ varies, the diffusion coefficient displays 
a maximum for $F\simeq F_3$ (see Fig. \ref{fig:Diff-sv}),
with a behavior similar to that found by
Reimann et al. in the IBP~\cite{Reimann}. As $\alpha$ decreases, however, 
we observe some differences.
 In particular, the larger diffusivity of the EPS
with respect to the IBP is basically explained by the excluded volume 
repulsion.
At first, the diffusivity lowers as $\alpha$ decreases, but
a further decrease of $\alpha$ determines a new enhancement of $D$ 
(Fig. \ref{fig:Diff-sv}). 
Such a feature is a consequence both of the larger broadening
and of the larger asymmetry 
of the packet occurring as the particle size grows 
(see Fig. \ref{fignx}); as a matter of fact,   
there is a huge increase of the diffusion since, 
for $F/F_3\simeq 0.88$, we found $D/D_0 \approx 29.6$
(these data are not shown).

\section{Underdamped regime}
When
the damping force, in Eq.~(\ref{kramers2}), is smaller or comparable 
with respect to the
inertial term $m\ddot{x}$ the situation changes drastically.

At low friction, the particles can travel across several wells
before being trapped in a minimum.  Again, as a guide, we can use
the results obtained by Risken in his thorough study 
on the IBP in washboard potentials,
providing a full theoretical treatment of the
mobility as a function of $\gamma$ and temperature.
Risken's theory shows the existence of two dynamical states:
{\it locked} and {\it running}.
At $T= 0$, a characterization of the 
dynamical behavior of the non interacting system is straightforward. Indeed, 
for $F>F_3=\omega_0^2$ only
running states exist. When $F_1\leq F\leq F_3$, with 
$F_1=4\gamma\pi\sqrt{m V_0}$ 
particles may be either locked or running depending on the
initial velocity and position. This fact determines a hysteresis 
loop in a $\mu$ vs $F$ diagram.
When the temperature is finite,  
the particles can switch from one state to the other
under the influence of the thermal bath.
As a result, the hysteresis is suppressed
and the locked-running transition occurs smoothly as a function
of the tilt $F$, being the $\mu(T)$ curve a sigmoid in $F$.
Only as $T\to 0^{+}$ the sigmoid becomes a step function whose discontinuity
is located at $F_2\simeq 3.36\gamma\sqrt{ m V_0}$. 

\begin{figure}[htb]
\includegraphics[clip=true,width=8.5cm, keepaspectratio,angle=0]
{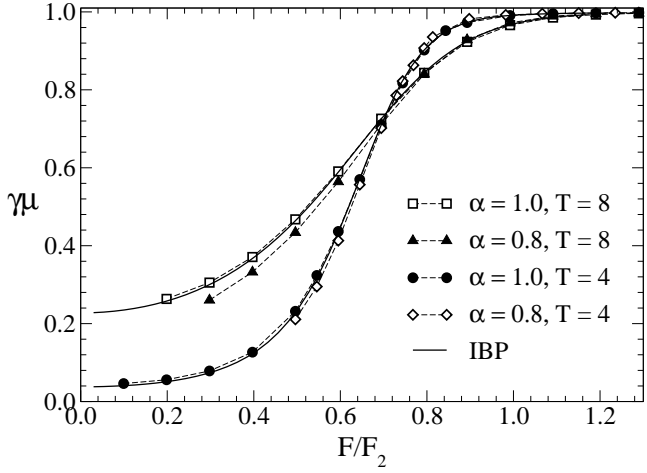}
\caption{Dimensionless mobility $\gamma \mu$ as a function of the
external rescaled force $F/F_2$ for different temperatures and coefficients of
restitution. The open and closed symbols refer to $T=4$ and
$T=8$, respectively. The squares indicate the elastic case and 
the triangles the case with $\alpha=0.8$. 
The other parameters of the system are 
$\gamma=0.2 t_u^{-1}$, $N=256$, $V_0=9.0$, $\sigma=1.0$, $w/\sigma=400$ and
$t_{max}=10000 t_u$. The full line indicates the corresponding curve 
for the non interacting
particles, obtained via the continued fraction method by Risken.}
\label{fig:mob-ud}
\end{figure}

MD simulations of the underdamped regime at different values of $F$ 
allowed us to determine the curve of $\mu$ 
as a function of the tilt, shown in Fig.~\ref{fig:mob-ud} for the EPS 
($\alpha=1$)  and the IPS with 
$\alpha=0.8$ at temperatures $T=8$ and $T=4$.

The mobility, $\mu$, appears to be roughly similar 
to that of the overdamped regime,
but the critical $F$ separating locked and unlocked 
situations lies at values lower than $F_3$ 
and depends both on $\gamma$ and $T$. 
In the EPS, the mobility is rather close to the IBP value,
while for the IPS the mobility is reduced, as shown explicitly
in the case  $\alpha=0.8$ in Fig.~\ref{fig:mob-ud}.

\begin{figure}[htb]
\includegraphics[clip=true,width=8.5cm,keepaspectratio,angle=0]
{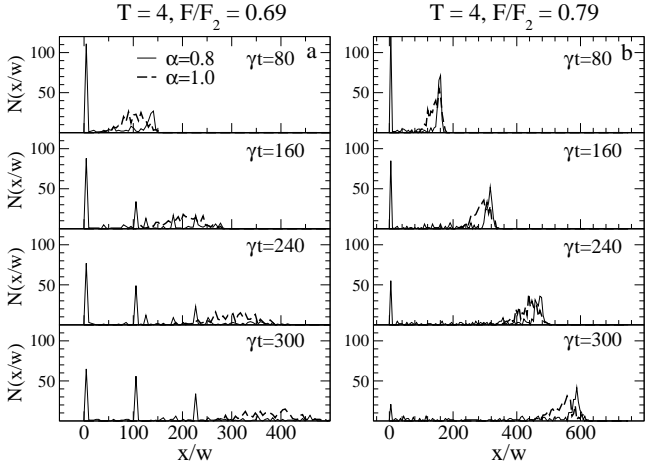}
\caption{Snapshots, taken a different times, of the averaged distribution 
$N(x,t)$ plotted as a function of the dimensionless quantity $x/w$, 
for $F=1.4$ (panel a) and $F=1.6$ (panel b), 
for the IPS with $\alpha=0.8$ and EPS (broken line). 
The remaining parameters are 
the same as in Fig.~\ref{fig:mob-ud}. 
We notice that the system 
size is $L=10^8 w$ much larger than the well width, thus the peaks are
not commensurate with the system size.}
\label{fig:N_xt_08}
\end{figure}

While, the mobility of particles interacting inelastically  
displays, in the underdamped regime, no peculiar behavior with 
respect to the IBP and EPS,
their collective diffusion presents some anomalies. 
At low temperatures, the dynamics strongly depends on the initial conditions, 
the evolution of the packet is very sensitive to the tilt and exhibits a 
rather inhomogeneous and irregular structure. 
This can be readily visualized by looking at the time behavior of the 
coarse grained particle density $N(x,t)$ [Eq.~\eqref{eq:rho-cg}] 
at temperature $T=4.0$ shown in  
figures ~\ref{fig:N_xt_08}(a) and ~\ref{fig:N_xt_08}(b) 
for $F=1.4$ and $F=1.6$, respectively. 
It indicates that, for $F=1.4$,
particles remain partially trapped in the well where they have been
initially deposited and only a fraction of them
escape acquiring a drift. 
For comparison we also show the corresponding behavior of 
$N(x,t)$ for the EPS, where the diffusion is
standard. Interestingly, the inelastic distribution $N(x,t)$
corresponds to a lower mobility, but to a larger spread of the particles.
Moreover, one can see that new clusters, indicated by the spikes
in Fig.~\ref{fig:N_xt_08}(a), spontaneously form and persist for long periods
before being dissolved. 
Such clustering phenomena are favored by the moderate value of the tilt.
The situation, instead, looks different at $F=1.6$
[see Fig.~\ref{fig:N_xt_08}(b)], where
the initial cluster ``evaporates'' earlier, and the
formation of new clusters is prevented by the effects of the drift $F$.
One observes a clear difference with
respect to the behavior of the overdamped IPS illustrated in Fig.
\ref{fignx} showing a more compact structure of $N(x,t)$ which
does not lose ``debris''.

\begin{figure}[htb]
\includegraphics[clip=true,width=8.0cm, keepaspectratio,angle=0]
{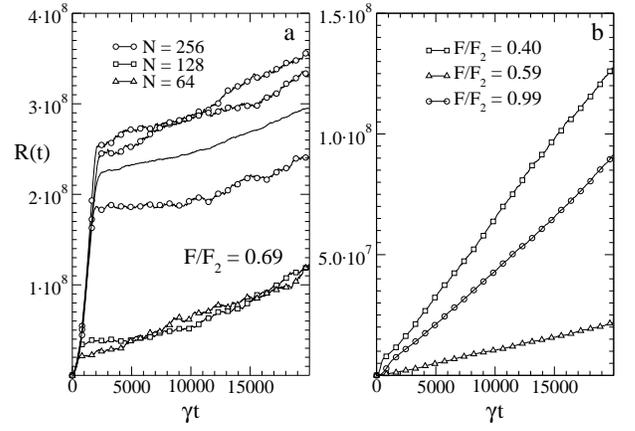}
\caption{Time behavior of $R(t)/\sigma^2$, \eqref{diffusion}, for the IPS with 
$\alpha=0.8$. Panel (a) reports three runs with $F=1.4$, $T=4$ 
and $N=256$ (circles) two runs with 
$N=128$ (squares) and two runs with $N=64$ (triangles). 
The full line is the average of the same quantity over 
$10$ independent runs involving $N=256$ particles.
Panel (b) shows $R(t)/\sigma^2$ vs $\gamma t$ at temperature $T=8$ 
for different forces: $F=0.8$ (circles), $F=1.2$ (squares) and $F=2.0$ 
(triangles). 
The curves represent an average over five runs. The remaining
parameters are the same as in Fig.\ref{fig:mob-ud}.}
\label{fig:sqrm}
\end{figure}

\begin{figure}[htb]
\includegraphics[clip=true,width=8.5cm, keepaspectratio,angle=0]
{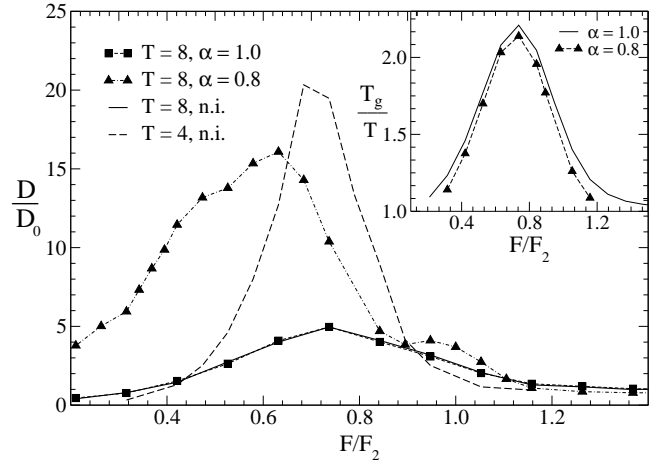}
\caption{Rescaled diffusion coefficient $D/D_0$ ($D_0 = T/m\gamma$), 
for different
coefficients of restitution, as a function of $F/F_2$. The $D$ values represent
averages over ten trajectories. The squares correspond to the
elastic case ($\alpha=1.0$), while the triangles to $\alpha=0.8$.
The temperature is $T=8.0$, the remaining parameters
are the same as in Fig.~\ref{fig:mob-ud}. 
 The dashed line indicates the result relative to the  IBP
at a lower temperature $T=4.0$, showing
that the effect of inelastic collisions cannot be accounted for by a mere
reduction of the effective temperature of the system. 
The inset, finally, displays the ratio 
between the kinetic and the heat bath temperatures.}
\label{fig:Diff-ud}
\end{figure}

Such an early stage is sufficient to determine a late  collective 
transport characterized by an anomalous spreading. The closely packed
initial configuration has deep
 repercussions on the late spreading, $R(t)$. This quantity, 
despite a very long simulation, does not appear to reach a linear 
or any other simple functional asymptotic 
dependence on time (Fig.~\ref{fig:sqrm}).  
The situation described above corresponds to 
an early very steep growth
of $R(t)$, growth that becomes slower 
after a characteristic time $\tau$ whose duration decreases with the tilt
amplitude. As shown in Fig.~\ref{fig:sqrm}, the vertical position
of the knee strongly depends on the number of particles.
By no means the $R(t)$ shows the expected  linear behavior of  
standard diffusion.
Such a feature is present only in the inelastic systems.
However, when the temperature is raised, even the IPS  
recovers a linear behavior as seen in Fig.~\ref{fig:sqrm}(b). 
The estimated diffusion
coefficient is shown in Fig.~\ref{fig:Diff-ud}. 

The peak of $D$ relative to the IPS is higher
than the corresponding peak of the EPS.
The larger value of the IPS peak can be associated with the
shape of the phase-space distribution, $P(x_r,v_r)$, which is defined
as the probability of finding a particle at distance $x_r$ from the
center of mass and with velocity $v_r$ with respect to center of mass
velocity $v_{CM}$.\\

\begin{figure}[htb]
\includegraphics[clip=true,width=9.0cm, 
keepaspectratio,angle=0]{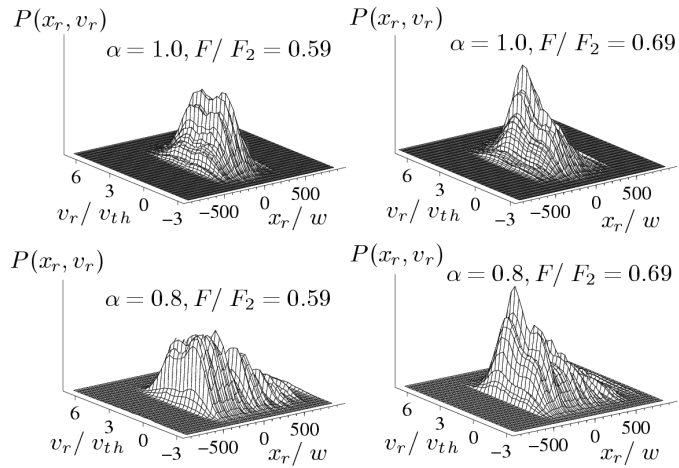}
\caption{Joint distribution $P(x_r,v_r)$ of position and velocity 
in the reference frame of system center of mass, plots refer to force 
$F=1.2$ (left) and $F=1.4$ (right) in the EPS case (top) and IPS case 
(bottom) with $\alpha=0.8$. Independent variables have been made 
dimensionless through the rescaling $x/x_{cm}$ and $v/v_{th}$, 
with $v_{th}=\sqrt{T/m}$ the thermal velocity.} 
\label{fig:sei}
\end{figure}

The enhancement of the diffusion in the IBP is determined by the 
locked-running bistability, i.e. by the existence of two 
peaks at velocities
$v=0$ and $v=F/\gamma$ in the velocity distribution. At the transition
both populations are present and the center of mass velocity 
does not represent the most probable velocity in the system. Hence, it 
is clear that in this situation the variance of the velocity distribution
can be very large. On the other hand, when one of the two populations
becomes dominant the variance tends to decrease.
The spatial part of the distribution $P(x_r,v_r)$ does not play any role
in the IBP as the $x$-dependence of the distribution is symmetric 
with respect to the center of mass coordinate. The scenario changes
in the interacting cases, since in one dimension, 
the excluded volume plays a fundamental
role in the dynamics. As we have seen in the inelastic case, at low 
temperatures the interaction was such as to lead to a fragmentation of the
system and to a non linear behavior of the diffusion.
As $T$ increases the fragmentation decreases and $R(t)$ display a linear
behavior with respect to $t$. 
In order to explain the different values of $D$ near the peak
between the IPS and the EPS, we have collected from the simulations
the double histograms $P(x_r,v_r)$.  
The comparison is reported in 
Fig. \ref{fig:sei} for drift values $F$ near the peak.
Analyzing the
different behaviors of $P(x_r,v_r)$
in the various cases may help clarify the role of the interactions.
In the EPS for loads $F=1.2$ and $1.4$, $P(x_r,v_r)$ displays the same
velocity bistability as the IBP and no spatial asymmetry and indeed 
the diffusion in these two systems results very similar as shown
in Fig. \ref{fig:Diff-ud}.
In the IPS, instead, for the same loads we observe a strong spatial
asymmetry which enhances the diffusion. In the IPS, the 
interactions besides determining the peak  of the 
diffusion also determine its width. From our simulations, as shown
in Fig.~\ref{fig:Diff-ud}, we see that the peak occurs at lower values 
of $F$ with respect
to the IBP and EPS and the width is much broader. This means that
it is not possible to reproduce the behavior of the IPS by an appropriate
choice of an effective temperature accounting for the inelasticity.
A smaller temperature, in fact, would give a higher peak, but
would not change the value of $F$ at which it occurs.
Moreover, the width of the peak decreases with $T$ and fails to reproduce
the observed broadening in the IPS, which is due to the inter-particle
interactions.
In the inset of Fig.~\ref{fig:Diff-ud}, we display the ratio between the
kinetic temperature 
$$
T_g=\frac{1}{N} \sum_i \langle (v_i-v_{CM})^2 \rangle
$$
and the heat bath temperature $T$.
Notice that in both cases the peak occurs in correspondence of the peak
of the curve $D/D_0$ of the elastic system and the ratio approaches, as 
expected, the 
value $1$ at low and high bias $F$. 


\section{Conclusions}
In summary,
in this work we have reported a numerical study on the dynamics and 
the transport
of an array of identical particles in a inclined washboard potential.
We investigated, at several tilting $F$, the role of hard-core repulsive 
interactions among the particles on both the mobility and the spreading 
of the system with respect to its center of mass, indicators
that are customarily used to characterize collective transport is such 
systems. We compared the behavior of three models involving 
respectively:  
independent Brownian particles (IBP), elastic particles (EPS) and 
inelastic particles (IPS). 
We have found that, in the high friction regime, the transport behavior of the 
system with interacting particles is qualitatively similar to that
of the non interacting system (IBP). Basically, the interactions do not 
strongly affect  the mobility 
which remains close to the IBP value, while they  
modify the diffusivity that appears to be larger. 
This conclusion is intuitive, if we consider that $\mu$ is related 
the center mass velocity of the system, a quantity rather insensitive to the 
presence of interparticle interactions. To achieve a 
a finer information about the structure of the collective motion, we
monitored the evolution of the particle spatial distribution. 
Such a distribution strongly depends on the choice of the geometrical 
parameters and of inelasticity. This analysis shows that 
an initial localized packet of particle spreads differently 
in the inelastic system from the elastic one.

The scenario in the underdamped regime is more complex.
The EPS systems, in fact, are similar to IBP either analyzing the 
mobility, or the diffusion of the particles. 
The inelastic interactions practically do not affect the 
mobility, whereas determine a marked change in the behavior of the 
diffusion at low temperatures and forces. 
The spreading, $R(t)$  (Eq. (\ref{diffusion})), 
presents a clear two stage behavior: it increases nonlinearly in a 
transient regime which is then followed by a nonlinear growth, 
as shown in Fig.~\ref{fig:sqrm}.
At higher temperatures, however, the diffusion
of the inelastic particles exhibits a single regime, 
with a linear growth typical of a standard diffusion. 
Our study indicates, moreover, that a group of
particles interacting via inelastic hard-core collisions
spread over much more than elastic particles. 

The mobility $\mu$ seems to be a global indicator that poorly 
encodes the information about the details of the interactions. 
On the contrary, the diffusion coefficient $D$ is a much more
representative observable being more sensitive to the presence 
of elastic or inelastic interactions.  
 
We believe that in the overdamped regime, it should be possible to apply
theoretical methods already employed in the study of dense 
molecular fluid in narrow channels~\cite{pedro}, whereas the
treatment of the underdamped case remains more problematic. 
The aim of the present numerical work 
is, perhaps, to stimulate further investigation and new theoretical proposals.

\acknowledgements
U.M.B.M. and F.C. acknowledge the support of the
Project Complex Systems and Many-Body Problems,
COFIN-MIUR 2005,2005027808.



\begin{thebibliography}{0}
\bibitem{polimeri}
G.I.~Nixon and G.W.~Slater, Phys.~Rev.~E {\bf 53}, 4969 (1996).

\bibitem{Nori1}
C.J.~Olson, C.~Reichhardt, B.~Janko, and F.~Nori, 
Phys.~Rev.~Lett. {\bf 87}, 177002 (2001).

\bibitem{Barone}
G.~Barone and A.~Patern\`o, {\it Physics and Applications of the Josephson
effect} (Wiley, New York, 1982).\\
G.~Barone and A.~Patern\`o, Phys.~Rev.~Lett. {\bf 87}, 177002 (2001).

\bibitem{adsorbate}
E.~Hershkowitz, P.~Talkner, E.~Pollak, and Y.~Georgievskii
Surf.~Sci. {\bf 421}, 73 (1999).

\bibitem{superionic} 
W.~Dieterich, P.~Fulde and I.~Peschel, 
Adv.~Phys. {\bf 29}, 527 (1980).

\bibitem{Prost}
F.~J\"ulicher, A.~Ajdari, and J.~Prost,
Rev. Mod. Phys. {\bf 69}, 1269 (1997).

\bibitem{Vicsek}
I.~Derenyi and T.~Vicsek, Phys.~Rev.~Lett. {\bf 75}, 374 (1995).

\bibitem{vulpi}
U. Marini Bettolo Marconi, M.~Conti and A.~Vulpiani,
Europhys.~Lett.~{\bf 51}, 685 (2000).

\bibitem{wolf}
S.~Dippel, G.G.~Batrouni and D.E.~Wolf,
Phys.~Rev.~E {\bf 54}, 6845 (1996).

\bibitem{Henrique}
C.~Henrique, M.A.~Aguirre, A.~Calvo, I.~Ippolito, S.~Dippel, G.G.~Batrouni, 
and D.~Bideau, Phys.~Rev.~E {\bf 57}, 4743 (1998).

\bibitem{Ancey} 
C.~Ancey, P.~Evesque, and P.~Coussot, 
J.~Phys.~(France) I {\bf 6}, 725 (1996). 

\bibitem{pipe}
T.~Poeschel, J.~Phys.~(France) I {\bf 4},  499 (1994).

\bibitem{Farkas}
Z.~Farkas, P.~Tegzes. A.~Vukics and and T.~Vicsek, 
Phys.~Rev.~E {\bf 60}, 7022 (1999).

\bibitem{Jaeger} 
H.M. Jaeger, S.R. Nagel and R.P. Behringer, 
Rev.~Mod.~Phys. {\bf 68}, 1259 (1996).

\bibitem{General1}
Granular Gas Dynamics, Lectures Notes in Physics vol. {\bf 624},
T.~P\"oschel and N.~Brilliantov editors, Berlin Heidelberg,
Springer-Verlag (2003) and references therein.

\bibitem{General3}
L.P.~Kadanoff, Rev.~Mod.~Phys. {\bf 71} 435 (1999).

\bibitem{General4}
J.~Duran {\em Sands, Powders and Grains. 
An Introduction to the Physics of Granular Materials},
(Springer, New York, 2000).

\bibitem{General5}
T.Poeschel and S.Luding (eds.)  {\em Granular Gases}, 
Lecture Notes in Physics vol. 564 (Berlin: Springer) (2001).

\bibitem{Paolotti}
D.~Paolotti, C.~Cattuto, U.~Marini-Bettolo-Marconi and A.~Puglisi
Granular Matter {\bf 5}, 75 (2003).

\bibitem{Risken}
H.~Risken, ``The Fokker-Planck Equation'' (Springer, Berlin, 1984).

\bibitem{Coffey}
W.T.~Coffey, Yu.P.~Kalmykov, S.V.~Titov and
B.P.~Mulligan, Phys.~Rev.~E {\bf 73}, 061101 (2006).

\bibitem{Reimann}
P.~Reimann C.~Van~den~Broeck, H.~Linke, P.~H\"anggi, 
J.M.~Rubi, and A.~P\'erez-Madrid, 
Phys.~Rev.~Lett. {\bf 87}, 010602 (2001).

\bibitem{Costantini}
G.~Costantini and M.~Marchesoni, Europhys.~Lett. {\bf 48}, 491 (1999).

\bibitem{Harris}
T.E.~Harris, J.~Appl.~Prob. {\bf 2}, 323 (1965).

\bibitem{Kollmann}
C.~Lutz, M.~Kollmann, and C.~Bechinger,
Phys. Rev. Lett. {\bf 93}, 026001 (2004).

\bibitem{Taloni} A.~Taloni and F.~Marchesoni,
Phys.~Rev.~Lett. {\bf 96}, 020601 (2004).

\bibitem{Biwell}
F.~Cecconi, A.~Puglisi, U.~Marini-Bettolo-Marconi and A.~Vulpiani,
Phys.~Rev.~Lett. {\bf 90} 064301 (2003).

\bibitem{CDBP}
F.~Cecconi, F.~Diotallevi, U.~Marini-Bettolo-Marconi, and A.~Puglisi,
J.~Chem.~Phys. {\bf 120}, 35 (2004) and J.~Chem.~Phys. {\bf 121}, 5125 (2004).

\bibitem{Pagnani}
R.~Pagnani, U.~Marini-Bettolo-Marconi, and A.~Puglisi
Phys.~Rev.~E {\bf 66}, 051304 (2002).

\bibitem{Kadanoff}
Y.~Du, H.~Li, and L.P.~Kadanoff, Phys.~Rev.~Lett. {\bf 74}, 1268
(1995).

\bibitem{Sela} 
N.~Sela and I.~Goldhirsch, Phys.~Fluids {\bf 7}, 507 (1995).

\bibitem{Mackintosh} 
D.R.M.~Williams and F.C.~MacKintosh, Phys.~Rev.~E {\bf 54}, R9
(1996).

\bibitem{Mcnamara} 
S.~McNamara and W.R.~Young, Phys.~Fluids~A {\bf 4}, 496 (1992);
{\em ibid.} {\bf 5}, 34 (1993).

\bibitem{Borromeo}
M.Borromeo and M.Marchesoni, Phys.Lett. A {\bf 249}, 8457 (1998).

\bibitem{Borromeo2}
M.~Marchesoni and M.~Borromeo, Phys.Rev.B {\bf 65}, 184101 (2002).

\bibitem{Stokes}
G.G.~Stokes, Trans. Cambridge Philos. Soc. {\bf 8}, 441 (1847).

\bibitem{Stokes2}
C.~Kettner, P.~Reimann, P.~H\"anggi, and F.~Muller
Phys.~Rev.~E {\bf 61} , 312 (2000)  

\bibitem{Lohse2}
D.~van~der~Meer, K.~van~der~Weele, and D.~Lohse,
Phys.~Rev.~E {\bf 63}, 061304 (2001)

\bibitem{Conti}
U.~Marini-Bettolo-Marconi and M.~Conti,
Phys. Rev. E {\bf 69} , 011302 (2004).

\bibitem{Numericalrecipes}
W.H.~Press, S.A.~Teukolsky, W.I.~Vetterling and B.P.~Flannery 
{\em Numerical Recipes in Fortran 77}, 
(Cambridge University Press, Cambridge, UK, 1992).

\bibitem{pedro}
U.~Marini-Bettolo-Marconi and P.~Tarazona, J.~Chem.~Phys., 
{\bf 124} 164901 (2006).

\end{thebibliography}
\end{document}